\documentclass[onefignum,onetabnum]{siamonline250211}

\usepackage{lipsum}
\usepackage{amsfonts}
\usepackage{graphicx}
\usepackage{epstopdf}
\usepackage{algorithmic}
\ifpdf
  \DeclareGraphicsExtensions{.eps,.pdf,.png,.jpg}
\else
  \DeclareGraphicsExtensions{.eps}
\fi

\usepackage{enumitem}
\setlist[enumerate]{leftmargin=.5in}
\setlist[itemize]{leftmargin=.5in}

\newsiamremark{remark}{Remark}
\newsiamremark{hypothesis}{Hypothesis}
\crefname{hypothesis}{Hypothesis}{Hypotheses}
\newsiamthm{claim}{Claim}
\newsiamremark{fact}{Fact}
\crefname{fact}{Fact}{Facts}

\usepackage{amsmath,amssymb}
\usepackage{subcaption}
\usepackage{xcolor} % need [dvipsnames]?
\definecolor{mygreen}{RGB}{134,180, 37}
\definecolor{myred}{RGB}{217, 83, 25}
\definecolor{mypurple}{RGB}{126, 47,142}
\definecolor{myyellow}{RGB}{237,177, 32}

\headers{Hamiltonian for active spheroidal particle motion}{B. Harding, R. N. Valani, and Y. M. Stokes}

\title{A Hamiltonian formulation for the motion of an active spheroidal particle suspended in laminar straight duct flow\thanks{Submitted to the editors \today.
\funding{This research is supported under the Australian Research Council’s Discovery Projects funding scheme (project number DP200100834) and the Royal Society of New Zealand's Marsden Fund (project ID 23-VUW-062). 
R.V. acknowledges the support of the Leverhulme Trust [Grant No. LIP-2020-014] and the ERC Advanced Grant ActBio (funded as UKRI Frontier Research Grant EP/Y033981/1).}}}

\author{Brendan Harding\thanks{School of Mathematics and Statistics, Victoria University of Wellington, Wellington, New Zealand 
  (\email{brendan.harding@vuw.ac.nz}, \url{https://people.wgtn.ac.nz/brendan.harding}).}
\and Rahil N. Valani\thanks{Rudolf Peierls Centre for Theoretical Physics, University of Oxford, OX1 3PU, United Kingdom
  (\email{rahil.valani@physics.ox.ac.uk}).}
\and Yvonne M. Stokes\thanks{School of Computer and Mathematical Sciences, The University of Adelaide, South Australia, Australia
 (\email{yvonne.stokes@adelaide.edu.au}).}}

\usepackage{amsopn}

\ifpdf
\hypersetup{
  pdftitle={Hamiltonian for active spheroidal particle motion},
  pdfauthor={B. Harding, R. N. Valani, and Y. M. Stokes}
}
\fi

\begin{document}

\maketitle

% REQUIRED
\begin{abstract}
We analyse a generalisation of Z\"{o}ttl and Stark's model of active spherical particles [Phys. Rev. Lett. 108, 218104 (2012)] and prolate spheroidal particles [Eur. Phys. J. E 36(1), 4 (2013)] suspended in cylindrical Poiseuille flow, to particle dynamics in an arbitrary unidirectional steady laminar flow through a straight duct geometry. 
Our primary contribution is to describe a Hamiltonian formulation of these systems and provide explicit forms of the constants of motions in terms of the arbitrary fluid velocity field.
The Hamiltonian formulation provides a convenient and robust approach to the computation of particle orbits whilst also providing new insights into the dynamics, specifically the way in which orbits are trapped within basins defined by a potential well.
In addition to considering spherical and prolate spheroidal particles, we also illustrate that the model can be adapted to oblate spheroidal particles.
\end{abstract}

% REQUIRED
\begin{keywords}
active particle, 
hamiltonian mechanics, 
spheroidal particle, 
Poiseuille flow, 
constant of motion
\end{keywords}

% REQUIRED
\begin{MSCcodes}
37N10, % Dynamical systems in fluid mechanics, oceanography and meteorology
37N25, % Dynamical systems in biology
70H99, % Hamiltonian and Lagrangian mechanics
76T20 % suspensions
\end{MSCcodes}

\section{Introduction}

Active particles is a generic term for particles, animate or inanimate, that consume energy and convert it into some form of self-propulsion~\cite{RevModPhys.88.045006}.
In the natural world, microscopic organisms such as bacteria can self-propel themselves via a variety of propulsion mechanisms with their moving appendages and deforming bodies, such as via the periodic beating of a flagellum (hairlike structures protruding from the organism)~\cite{Pismen2021_Ch4}.
In the realm of the human-made world, examples include active colloids such as Janus particles that self-propel by breaking fore-aft symmetry via asymmetric coating of surface properties in an appropriate environment~\cite{Pismen2021_Ch3}.
Both settings have sparked much interest, both in terms of trying to understand the mechanisms underlying these fascinating phenomena and attempts to exploit active particle motion for a wide range of practical applications.
Active particle motion typically occurs within a fluid film or suspension~\cite{doi:10.1146/annurev-chembioeng-060817-084006}.
While many studies focus on active particle motion within a stationary fluid~\cite{Lauga_2009,PhysRevFluids.9.083302}, there has been a surge of interest in understanding active particle motion in flowing fluids~\cite{RUSCONI20151,Microswimmersreview}.
Many applications invariably involve transporting of active particles from one place to another through pipes/ducts, such as sperm cells swimming in Fallopian tubes~\cite{doi:10.1073/pnas.1120955109,Simons2018}, pathogens moving through blood vessels~\cite{doi:10.1146/annurev-chembioeng-060817-084006} and micro-robots programmed for targeted drug delivery applications~\cite{D0CS00309C}, so this is a natural flow setup to begin with.

Z\"ottl and Stark introduced and studied a model of a simple active spherical particle suspended in Poiseuille flow through a cylindrical pipe \cite{ZottlStark2012}.
The model consists of an idealised active particle which is spherical, infinitesimal in size (i.e. not influencing the fluid motion), swims at a constant velocity, and swims in a direction which is only modified via the rotation of the active particle due to gradients in the fluid velocity field.
Subsequently, they considered the case of prolate spheroidal particles in which the axis aligned with the swimming direction has length $\gamma>1$ relative to two perpendicular axes \cite{ZottlStark2013}. 
In both works they utilised a parametrisation of the direction vector $\mathbf{e}$ in spherical coordinates.
In the case of spherical active particles they showed the essential dynamics can be described by a system of differential equations involving four variables which possesses two constants of motion.
In the case of prolate spheroidal particles they similarly showed the essential dynamics can be described by a system of differential equations involving four variables and identified one constant of motion. However, the quasi-periodic nature of their results clearly indicated there was likely a second unidentified constant of motion (since motion confined to a two-dimensional manifold satisfies the requirements of the Poincare--Bendixson theorem and thereby rules out the possibility of chaotic orbits).

In a recent paper we examined the application of Z\"ottl and Stark's spherical active particle model in the context of square duct flow \cite{ValaniHardingStokes2024}.
In the absence of rotational symmetry of the flow field, there is only one constant of motion associated with the coupled system of four differential equations.
As such, this opens up the possibility of chaotic orbits and thereby greatly increases the variety in the dynamics observed.
We devised a classification of the many different types of orbits observed in the system and applied it in a thorough analysis of the space of initial conditions.

In this paper we make several contributions to the active spherical and spheroidal particle models of Z\"ottl and Stark~\cite{ZottlStark2012,ZottlStark2013}.
We generalise the models to (unidirectional) steady laminar straight duct flows with arbitrary cross-section shape.
We show that a parametrisation of the direction vector in spherical coordinates can be avoided and the equations reformulated in a manner which leads to a Hamiltonian.
We examine the case of $0<\gamma<1$ in addition to the cases $\gamma=0$ and $\gamma>1$, corresponding to oblate spheroidal, spherical and prolate spheroidal particles, respectively.

\section{The active particle model}

Consider a straight duct aligned with the $z$-axis through which there is a steady laminar fluid flow of the form 
$$\mathbf{u}_f(x,y)=w(x,y)\mathbf{k}\,,$$
with $\mathbf{i},\mathbf{j},\mathbf{k}$ being the standard Cartesian basis of $\mathbb{R}^3$.
While the development of the model of active particle motion places little restriction on the specific nature of $w(x,y)$, other than it being continuously differentiable, it is convenient to consider the flow to be the solution, or approximate solution, of a Poisson problem with Dirichlet boundary conditions.
Specifically, we consider $w(x,y)$ to (approximately) solve
\begin{equation}\label{eqn:poisson}
\nabla^2w=-P \quad\text{in $\mathcal{C}$,} \quad w=0 \quad\text{on $\partial\mathcal{C}$,}
\end{equation}
where $P>0$ is a constant denoting the pressure gradient driving the flow, $\mathcal{C}\subset R^2$ is a bounded and connected set describing the cross-section of the duct, i.e. so that the fluid domain is $\{(x,y,z)\in\mathbb{R}^3:(x,y)\in \mathcal{C}\}$, and $\partial\mathcal{C}$ denotes the cross-section boundary.
In this context, the flow is unidirectional (i.e. $w\geq 0$) and the critical points of $w$ are necessarily (local) maxima or saddle points.
For most common/practical cross-sections there is a single (and thus global) maxima, although complex cross-sections can produce several local maxima and saddle points (for example, as is the case for the cross-section of \cref{fig:example_with_saddle}).

\begin{figure}
\centering
\includegraphics[width=0.5\textwidth]{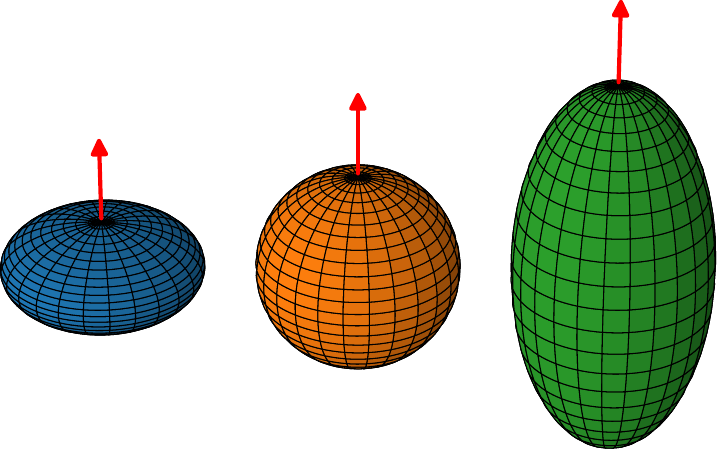}
\caption{Examples of the spheroidal particle shapes considered herein. 
From left to right we have: an oblate spheroidal particle ($\gamma=0.5$); a spherical particle ($\gamma=1$); and a prolate spheroidal particle ($\gamma=2$). 
The red arrows illustrate the vector $\mathbf{e}$ denoting both the axis of elongation and the swimming direction.}\label{fig:particle_shapes}
\end{figure}

Let 
$$\mathbf{x}(t)=x(t)\mathbf{i}+y(t)\mathbf{j}+z(t)\mathbf{k}$$ 
denote the location of the centroid of an active spheroidal particle which is elongated by a factor $\gamma$ along the unit vector $\mathbf{e}(t)$ which also denotes the direction in which the particle swims (i.e. propels itself) with a constant velocity $U$.
Examples of such spheroidal particles are shown in \cref{fig:particle_shapes}.
The particle is assumed to be infinitesimal in size (so as not to disturb the fluid), non-inertial (so that it instantly responds to changes in the fluid velocity as its location in the cross-section changes) and does not interact directly with the walls of the duct which confine the flow. 

We non-dimensionalise the problem using the particle swimming velocity as the velocity scale $U$, half of the duct height as the length scale $L$ and $T=L/U$ as the time scale. 
With these choices, it takes unit dimensionless time for the particle to swim a distance equal to half to the duct height.
Moreover, with $w(x,y)$ non-dimensionalised with respect to the velocity scale $U$, its magnitude should be interpreted as the fluid velocity relative to the swimming velocity of the active particle. 
That is, 
\begin{equation}
W:=\max_{(x,y)\in\mathcal{C}}\{w(x,y)\} \label{eqn:max_fluid_vel}
\end{equation}
can be thought of as a parameter (proportional to the pressure gradient $P$) which adjusts the relative magnitude of the fluid velocity to the particle swimming velocity.

%%% Addition
For results utilising a square cross-section, $\mathcal{C}=[-1,1]^2$, we use the series solution for Hagen--Poiseuille flow, specifically
\begin{equation}\label{eqn:square_poiseuille}
w(x,y) = W\left(1-y^2+4\sum_{n=1}^{\infty}\frac{(-1)^{n}\cosh(k_n x)\cos(k_n y)}{k_n^3\cosh(k_n)}\right)\bigg/\left(1+4\sum_{n=1}^{\infty}\frac{(-1)^{n}}{k_n^3\cosh(k_n)}\right)\,,
\end{equation}
where $k_n:=(2n-1)\pi/2$ and the sums are truncated beyond $n=10$.
For results using a circular cross-section $\mathcal{C}=\{(x,y):x^2+y^2\leq 1\}$, we use the solution
\begin{equation}\label{eqn:circle_poiseuille}
w(x,y) = W\left(1-x^2-y^2\right)\,.
\end{equation}

Following \cite{ZottlStark2013}, the motion of the particle is governed by the coupled (dimensionless) differential equations
\begin{subequations}\label{eqn:model_compact_form}
\begin{align}
\frac{d\mathbf{x}}{dt}&=\mathbf{e}+w(x,y)\mathbf{k} \,, \\
\frac{d\mathbf{e}}{dt}&=\boldsymbol{\Omega}\times \mathbf{e} \,,
\end{align}
\end{subequations}
where
$$\boldsymbol{\Omega}=\frac{1}{2}\boldsymbol{\Omega}_f+G\mathbf{e}\times[\mathbf{E}\cdot\mathbf{e}]\,,$$
with $G=(\gamma^2-1)/(\gamma^2+1)$ being a geometry factor,
$$\boldsymbol{\Omega}_f=\nabla\times\mathbf{u}_f=\nabla\times(w(x,y)\mathbf{k})=\frac{\partial w}{\partial y}\mathbf{i}-\frac{\partial w}{\partial x}\mathbf{j}\,,$$
and
$$\mathbf{E}=\frac{1}{2}\left[\nabla\mathbf{u}_f+(\nabla\mathbf{u}_f)^\intercal\right]=
\frac{1}{2}\begin{bmatrix}0&0&\frac{\partial w}{\partial x}\\
0&0&\frac{\partial w}{\partial y}\bigg.\\ % Note: the \bigg. adds some space between rows
\frac{\partial w}{\partial x}&\frac{\partial w}{\partial y}&0\end{bmatrix}\,.$$

In expanded component form, using subscripts to denote the components of $\mathbf{e}$, the equations of motion read 
\begin{subequations}\label{eqn:model_component_form}
\newline\noindent\begin{minipage}{0.33\textwidth}
\begin{align}
\dot{x}&= e_x \,, \bigg.\\
\dot{y}&= e_y \,, \bigg.\\
\dot{z}&= e_z+w(x,y) \,, \bigg.
\end{align}
\end{minipage}\hfill
\begin{minipage}{0.66\textwidth}
\begin{align}
\dot{e}_x &= -\frac{1}{2}(1-G(1-2e_x^2))e_z\frac{\partial w}{\partial x}-Ge_xe_ye_z\frac{\partial w}{\partial y} \,, \bigg.\\
\dot{e}_y &= -\frac{1}{2}(1-G(1-2e_y^2))e_z\frac{\partial w}{\partial y}-Ge_xe_ye_z\frac{\partial w}{\partial x} \,, \bigg.\\
\dot{e}_z &=  \frac{1}{2}(1+G(1-2e_z^2))\left(e_x\frac{\partial w}{\partial x}+e_y\frac{\partial w}{\partial y}\right) \,. \bigg. \label{eqn:dotez_general}
\end{align}
\end{minipage}\vskip1em\noindent 
\end{subequations}
Notice that no equation is dependent on $z(t)$, and so its evolution decouples from the system of equations.
There is one universal constant of motion obtained from the fact that $\mathbf{e}$ should remain a unit vector, that is one has
\begin{equation}
\|\mathbf{e}(t)\|^2=e_x(t)^2+e_y(t)^2+e_z(t)^2=1 \,.\label{eqn:unit_e}
\end{equation}
Z\"ottl and Stark implicitly utilise this when parametrising $\mathbf{e}$ in terms of the two angular coordinates of a spherical coordinate system~\cite{ZottlStark2012,ZottlStark2013}.
We will avoid using such a parametrisation herein.

There is a second universal constant of motion associated with \cref{eqn:model_component_form}, although its precise form differs depending on the value of $G$ (as will be seen in subsequent sections).
As a consequence, the system generally exhibits the motions of a three dimensional (continuous) dynamical system.
In the special case where $w(x,y)$ possesses rotational symmetry, such as Poiseuille flow through a cylindrical pipe, there is one additional constant of motion which reduces the dynamics to that of a two dimensional (continuous) dynamical system (thus ruling out the possibility of chaotic orbits due to the Poincare--Bendixson Theorem).

In the subsequent sections we analyse this active particle model in the case of spherical ($G=0$), prolate spheroidal ($0<G<1$) and oblate spheroidal ($-1<G<0$) particles, in each case illustrating how the differential equations can be reformulated as being derived from a Hamiltonian system and providing the explicit form of constants of motion in terms of a general flow function $w(x,y)$.

Stationary points of \cref{eqn:model_component_form} (ignoring the evolution of $z(t)$) occur where $e_x=e_y=0$, implying $e_z=\pm1$, and $\partial w/\partial x=\partial w/\partial y=0$, i.e. corresponding to critical points of $w(x,y)$.
As discussed, the solution of \cref{eqn:poisson} for typical cross-section shapes results in a fluid velocity $w(x,y)$ possessing a unique (global) maximum.
At this maximum, a small deviation from the orientation $\mathbf{e}=(0,0,1)$ generally results in motion away from the stationary point whereas a small deviation from the orientation $\mathbf{e}=(0,0,-1)$ results in motion oscillatory motion around the stationary point.
With this observation, we generally view the orientation $\mathbf{e}=(0,0,1)$ as being unstable.
A thorough derivation and classification of all stationary points is undertaken for each case of active particle shape examined below.

\section{A Hamiltonian formulation of active spherical particle motion}\label{sec:spherical_hamiltonian}

In the special case $\gamma=1$ (and thus $G=0$) the particle has a spherical shape and the equations \cref{eqn:model_component_form} reduce to 
\begin{align*}
\dot{x}&={e}_x \,, &
\dot{e}_x &= -\frac{1}{2}e_z\frac{\partial w}{\partial x} \,, \\
\dot{y}&={e}_y \,, &
\dot{e}_y &= -\frac{1}{2}e_z\frac{\partial w}{\partial y} \,, \\
\dot{z}&={e}_z+w(x,y) \,, &
\dot{e}_z &=  \frac{1}{2}\left(e_x\frac{\partial w}{\partial x}+e_y\frac{\partial w}{\partial y}\right) \,.
\end{align*}
This case was thoroughly explored in the case of cylindrical pipe flow by Z\"ottl and Stark \cite{ZottlStark2012} and in the case of square duct flow by Valani et. al \cite{ValaniHardingStokes2024}, but we briefly re-examine it here to make note of a Hamiltonian formulation of the system not previously identified.

Observe that a second constant of motion (i.e. additional to \cref{eqn:unit_e}) may be expressed as
\begin{equation}\label{eqn:CoM_Geq0}
e_z(t)-\frac{1}{2}w(x(t),y(t)) = C_s \,,
\end{equation}
where $C_s$ is determined by the initial values $x(0),y(0),e_z(0)$, i.e.
\begin{equation*}
C_s:=e_z(0)-\frac{1}{2}w(x(0),y(0)) \,.
\end{equation*}
Observe that \cref{eqn:CoM_Geq0} provides a precise relationship between the $(x,y)$ coordinates of a given orbit and the value of $e_z$.
Further, ignoring the evolution of $z(t)$, we can now describe the essential dynamics of the system via the second order equations
\begin{subequations}\label{eqn:2nd_order_model_Geq0}
\begin{align}
\ddot{x} &= \dot{e}_x = -\frac{1}{2}e_z\frac{\partial w}{\partial x} = -\frac{1}{2}\left(C_s+\frac{1}{2}w(x,y)\right)\frac{\partial w}{\partial x} \,, \\
\ddot{y} &= \dot{e}_y = -\frac{1}{2}e_z\frac{\partial w}{\partial y} = -\frac{1}{2}\left(C_s+\frac{1}{2}w(x,y)\right)\frac{\partial w}{\partial y} \,.
\end{align}
\end{subequations}
By observation, the right-hand sides of the equations \cref{eqn:2nd_order_model_Geq0} can be represented as the negation of the gradient of a scalar potential function
\begin{equation*}V_s(x,y)=\frac{1}{2}\left(C_s+\frac{1}{2}w(x,y)\right)^2\,.\end{equation*}
Using the nomenclature of Hamiltonian mechanics, we define the momenta $P_x=\dot{x}$ and $P_y=\dot{y}$ for which it is readily checked that 
\begin{equation}
\mathcal{H}=\frac{1}{2}(P_x^2+P_y^2)+V_s(x,y) \,,\label{eqn:Hamiltonian_Geq0}
\end{equation}
is a Hamiltonian which describes the dynamics of the system (that is the second order equations \cref{eqn:2nd_order_model_Geq0} are recovered via Hamilton's equations applied to \cref{eqn:Hamiltonian_Geq0}).

\begin{figure}
\begin{center}
\includegraphics{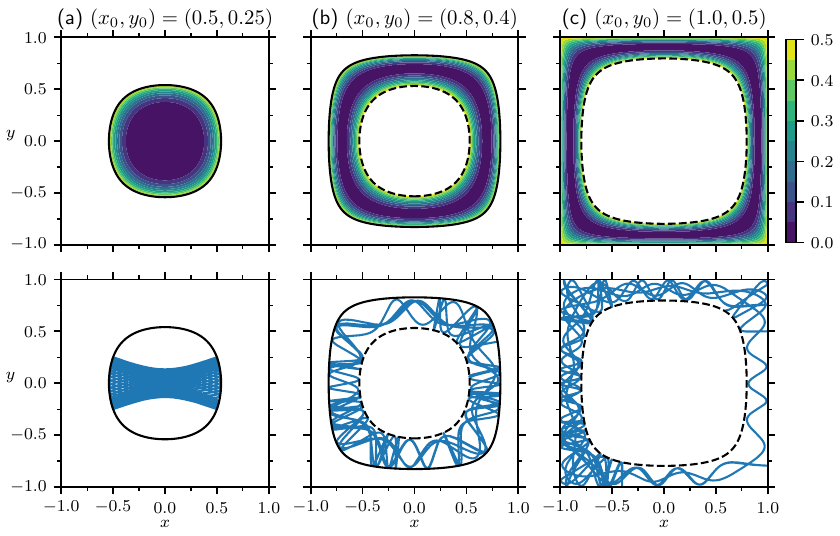}
\end{center}
\caption{Several examples of potentials and (cross-sectional) orbits of an active spherical particle in square duct flow, specifically utilising \cref{eqn:square_poiseuille}. 
The top row shows the boundaries of the the potential $V_s$ within the basin $V_s\leq0.5$ for different values of $C_s$ as determined from the stated initial conditions for $(x_0,y_0)$ along with $\mathbf{e}_0=(0,0,-1)$.
Black solid and dashed lines show where $V_s=1/2$ corresponding to particle motion perpendicular to the cross-section, specifically with $e_z=-1$ and $e_z=+1$, respectively.
The bottom row illustrates the orbit corresponding to the stated initial conditions.
The left column illustrates an orbit whose basin is confined near the centre, the middle column illustrates an orbit whose basin is confined to an annular region, and the right column illustrates a basin which is confined to a neighbourhood of the duct wall.
Observe that the confinement is tight.
}\label{fig:basins_spherical_square}
\end{figure}

\begin{figure}
\begin{center}
\includegraphics{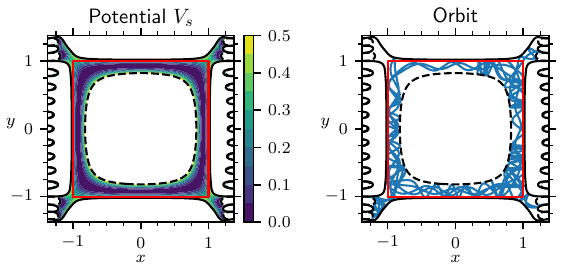}
\end{center}
\caption{An example of a potential and orbit of an active spherical particle in square duct flow, specifically utilising \cref{eqn:square_poiseuille}, which extends beyond the cross-section $\mathcal{C}=[-1,1]^2$ (boundary shown in red). 
The initial conditions are $(x_0,y_0)=(1,0.5)$, $\mathbf{e}_0=(0,0.6,-0.8)$.
Note that despite the orbit escaping the cross-section, it still remains confined to the region for which $V_s\in[0,1/2]$ (i.e. upon applying \cref{eqn:square_poiseuille} outside of the domain $\mathcal{C}$). 
The contours of $V_s=1/2$ (black solid and dashed lines) which lie outside of $\mathcal{C}$ arise from the truncation of the series \cref{eqn:square_poiseuille}.
Compare with the examples in \cref{fig:basins_spherical_square}.
In a physical/experimental setting we would expect the particle to remain adjacent to the wall until the direction vector has turned toward the interior of the cross-section.
}\label{fig:basins_spherical_square_escape}
\end{figure}

\begin{figure}
\centering
\includegraphics{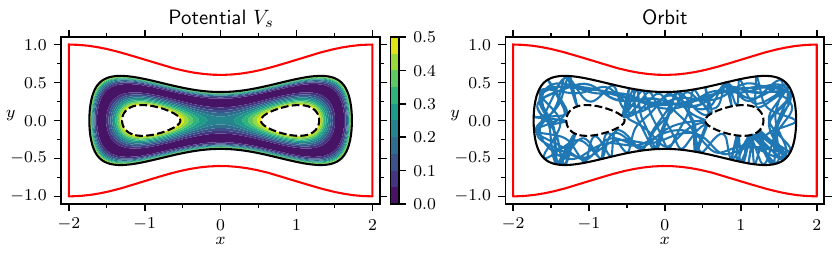} 
\caption{The potential $V_s$ (left) corresponding to an orbit (right) of a spherical active particle within flow through a cross-section for which the velocity field features a saddle.
The cross-section is outlined in red with top wall following the curve $y=1-0.4\cos(\pi x/4)^2$, which is reflected about $y=0$ to obtain the bottom wall.
The flow $w(x,y)$ is the solution of \cref{eqn:poisson} with $P\approx 41.19$ such that $W=\max\{w\}=10$.
The solution was approximated using the finite element method with roughly 32,000 quadratic elements.
The specific initial condition for the spherical active particle is $\mathbf{e}_0=(0,0,-1)$ and $(x_0,y_0)=(0.8,0.5)$.
}\label{fig:example_with_saddle}
\end{figure}

This formulation of the problem has several benefits.
By avoiding a parametrisation of the direction vector $\mathbf{e}$ in spherical coordinates we circumvent issues that potentially occur with singularities of trigonometric functions such as $\tan()$ at specific orientations.
In addition, while the second order formulation makes it possible to employ highly efficient integrators for second order problems, the
alternative Hamiltonian formulation makes it possible to employ symplectic integrators to ensure that constants of motion do not drift significantly from their initial values over long integration times.
Note that in our formulation of the Hamiltonian we have explicitly utilised the constant of motion \cref{eqn:CoM_Geq0}, such that it will always be satisfied within floating point error, and it can be observed that the Hamiltonian \cref{eqn:Hamiltonian_Geq0} is equivalently expressed as
$$\mathcal{H}=\frac{1}{2}(e_x^2+e_y^2+e_z^2)\,,$$
so that its preservation will maintain the unit length of $\mathbf{e}$.
Lastly, the potential $V_s(x,y)$ provides insight into why orbits remain trapped within specific regions of the cross-section, specifically those for which $V_s(x,y)\le1/2$. 
Noting that $V_s(x,y)=e_z^2/2$, % and how the switching of direction of motion around the origin can occur.
%%% BH Note: I use subscript "s" on V here to denote the "s"pherical version of the potential. I'll later use "p" and "o" for the prolate and oblate versions. 
the contour $V_s(x,y)=0$ describes locations where all of the ``energy'' is stored in the kinetic component of the Hamiltonian (i.e. $e_z=0$ and $\dot{x}^2+\dot{y}^2=e_x^2+e_y^2=1$, so that the swimming component of particle motion is entirely within a cross-section of the duct), whereas contours $V_s(x,y)=1/2$ describe locations where all of the energy is stored in the potential component (i.e. $e_z=\pm 1$,  $\dot{x}^2+\dot{y}^2=e_x^2+e_y^2=0$, and the motion of the particle is parallel to the direction of fluid flow).
With $V_s=1/2$ and $e_z=+1$, observe that $\ddot{x}=-\tfrac{1}{2}\partial w/\partial x$ and $\ddot{y}=-\tfrac{1}{2}\partial w/\partial y$ so that, provided $(x,y)$ is not a critical point of $w$, subsequent motion is in the direction $-\nabla w$.
Analogously, with $V_s=1/2$ and $e_z=-1$ then $\ddot{x}=\tfrac{1}{2}\partial w/\partial x$ and $\ddot{y}=\tfrac{1}{2}\partial w/\partial y$ so that, provided $(x,y)$ is not a critical point of $w$, subsequent motion is in the direction $\nabla w$.
Consequently, in the typical situation where $w$ has a unique maximum, the segment of the contour $V_s=1/2$ for which $e_z=1$ is an inner boundary of the basin whilst the segment for which $e_z=-1$ is an outer boundary of the basin.
See \cref{fig:basins_spherical_square,fig:basins_spherical_square_escape} for several examples involving a square cross-section.
An example involving a cross-section in which $w(x,y)$ possesses a saddle point is also provided in \cref{fig:example_with_saddle}.
It can be observed that one of the example orbits is quasi-periodic while the others have chaotic qualities to them.

%%% New example

One immediate application of these observations is that we can rigorously distinguish between ``tumbling'' and ``swinging'' (or ``wandering'') motion. 
Assume a typical flow in which $w$ has a unique maximum, with $W=\max\{w\}$ as defined in \cref{eqn:max_fluid_vel}. 
Then ``tumbling'' motion may be defined as any motion for which $e_z=1$ is achieved away from the position at which $w=W$ since this results in the particle turning away from the maximum and tumbling between the two contours $V_s=1/2$ corresponding to $e_z=1$ and $e_z=-1$. 
Moreover, since $e_z=C_s+w/2\leq C_s+W/2$, tumbling motion cannot occur where the initial conditions are such that $C_s<1-W/2$. 
Conversely, when the basin interior $V_s(x,y)=e_z^2/2<1/2$ contains the point of maximum $w$ at which $e_z=C_s+W/2<1$ (and thus $C_s<1-W/2$), then $e_z=1$ cannot be achieved for any initial conditions having this basin (as there is no contour $V_s=1/2$ on which $e_z=1$), so tumbling motion cannot occur. 

In all cases that $C_s<1-W/2$, and tumbling motion is not possible, we instead obtain a motion that we describe as ``swinging'' or ``wandering''.
We suggest the essential distinction between swinging and wandering motion be that swinging is either periodic or quasi-periodic (with respect to its limit set), whereas wandering is chaotic. 
One might go a little further and distinguish two sub-types of swinging/wandering motions. 
The first is for $C_s\le -W/2$ so that $e_z=C_s+w/2\le 0$ and $V_s$ is minimised at the point of maximum $w$. 
We call this ``simple swinging/wandering'' because the particle orbits have the property that no component of the particle motion, i.e. $\mathbf{e}$, is ever directed downstream. 
The second type, which we call ``complex swinging/wandering'' is for $-W/2<C_s<1-W/2$ so that $e_z$ can attain both negative and positive values and the potential $V_s$ possesses a local maximum at the point of maximum $w$.

The same classification of tumbling and swinging/wandering active particle motions can be applied to flows $w(x,y)$ in which the maximum value $w=W$ is achieved at multiple locations, as in \cref{fig:example_with_saddle}. 
The only difference is that, when it exists, the contour corresponding to $V_s=1/2,\ e_z=1$ may consist of distinct curves around each of the positions at which $w$ is maximised.

\subsection{General classification of stationary points for spherical active particles.}

Consider the stationary points of \cref{eqn:2nd_order_model_Geq0}.
Their location and properties can generally be inferred from the potential $V_s$ but, given the interplay between the initial orientation of the particle and the specific nature of $V_s$, we find it more straightforward to examine \cref{eqn:2nd_order_model_Geq0} as a first order system, specifically
\begin{align*}
\dot{x}&=P_x \,, & \dot{P_x}&=-\frac{1}{2}\left(C_s+\frac{1}{2}w(x,y)\right)\frac{\partial w}{\partial x} \,, \\
\dot{y}&=P_y \,, & \dot{P_y}&=-\frac{1}{2}\left(C_s+\frac{1}{2}w(x,y)\right)\frac{\partial w}{\partial y} \,.
\end{align*}

Noting that $P_x=e_x$ and $P_y=e_y$, then the requirement $\dot{x}=\dot{y}=0$ implies $C_s+w(x,y)/2=e_z=\pm 1$.
Consequently, the requirement $\dot{P}_x=\dot{P}_y=0$ implies $\partial w/\partial x = \partial w/\partial y=0$. 
Therefore, stationary points can only occur at critical points of $w(x,y)$.

The Jacobian of the first order system above has the general form
$$
\begin{bmatrix}0&0&1&0\\0&0&0&1\\a&b&0&0\\b&c&0&0\end{bmatrix}\,, 
\quad\text{where}\quad
\begin{array}{l} 
a = -\frac{1}{2}\left(C_s+\frac{1}{2}w(x,y)\right)\frac{\partial^2 w}{\partial x^2}(x,y)-\frac{1}{4}\left(\frac{\partial w}{\partial x}(x,y)\right)^2 \,,\\[2pt]
b = -\frac{1}{2}\left(C_s+\frac{1}{2}w(x,y)\right)\frac{\partial^2 w}{\partial x\partial y}(x,y)-\frac{1}{4}\frac{\partial w}{\partial x}(x,y)\frac{\partial w}{\partial y}(x,y) \,,\\[2pt]
c = -\frac{1}{2}\left(C_s+\frac{1}{2}w(x,y)\right)\frac{\partial^2 w}{\partial y^2}(x,y)-\frac{1}{4}\left(\frac{\partial w}{\partial y}(x,y)\right)^2 \,.
\end{array}
$$
At a stationary point, the coefficients $a,b,c$ reduce to 
\begin{equation*}
a = -\frac{1}{2}e_z\frac{\partial^2 w}{\partial x^2}(x,y) \,,\qquad
b = -\frac{1}{2}e_z\frac{\partial^2 w}{\partial x\partial y}(x,y) \,,\qquad
c = -\frac{1}{2}e_z\frac{\partial^2 w}{\partial y^2}(x,y) \,,
\end{equation*}
where we have further simplified using $e_z=C_s+w/2$, recalling one must have $e_z=\pm 1$.
In any case, the eigenvalues of the Jacobian satisfy
$$\lambda^2=\frac{a+c}{2}\pm\frac{1}{2}\sqrt{(a+c)^2-4(ac-b^2)}\,.$$
Notice that if $ac-b^2<0$ then the corresponding critical point of $w(x,y)$ is necessarily a saddle point and we obtain a pair of purely imaginary eigenvalues along with two real eigenvalues of opposite sign.
If $ac-b^2=0$ then the corresponding critical point of $w(x,y)$ is unstable but cannot be classified in detail without further analysis (as we obtain the eigenvalue zero with multiplicity two, along with two real eigenvalues of opposite sign).
Therefore, stationary points giving rise to either of the above two cases are unstable.
If $ac-b^2>0$ and $a,c<0$ then, as $(a+c)^2>(a+c)^2-4(ac-b^2)=(a-c)^2+4b^2\geq0$, we obtain four distinct purely imaginary eigenvalues.
If $ac-b^2>0$ and $a,c>0$ then we obtain four distinct real eigenvalues, two pairs having opposite sign.

Consider when $w$ solves the Poisson equation \cref{eqn:poisson}, that is $\nabla^2 w=-P$ for some constant $P>0$.
In the case that $ac-b^2<0$, the corresponding critical point of $w(x,y)$ is necessarily a saddle.
Analogously, in the case that $ac-b^2>0$, the corresponding critical point of $w(x,y)$ is necessarily a (local) maximum.
Further, as $a+c=\frac{1}{2}e_z P$, the two sub-cases of $ac-b^2>0$ are distinguished purely by the sign of $e_z$.
Specifically, $e_z=-1$ leads to the four purely imaginary eigenvalues meaning that small perturbations around this stationary point result in periodic orbits around it, i.e. the stationary point is a centre. 
On the other hand, $e_z=+1$ leads to the four distinct real eigenvalues, two positive and two negative, so that this stationary point behaves like a saddle point (and is therefore unstable).

\subsection{Special case of an active spherical particle in cylindrical pipe flow}

\begin{figure}
\begin{center}
\includegraphics{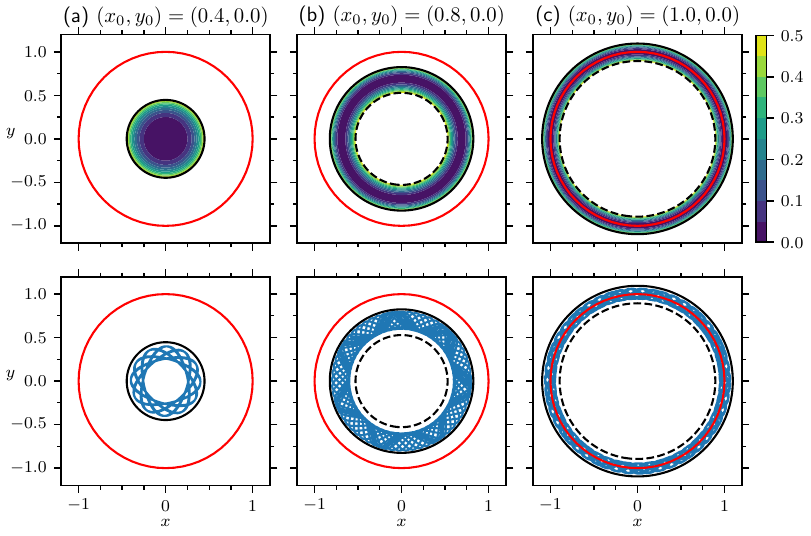}
\end{center}
\caption{Several examples of potentials and orbits of an active spherical particle in cylindrical pipe flow, specifically \cref{eqn:circle_poiseuille} with $W=10$. 
%The top row shows the boundaries of the basin (black solid and dashed lines corresponding to $e_z=-1$ and $e_z=+1$, respectively, as before) and the potential $V_s$ within the basin for different values of $C_s$ as determined from the stated initial conditions for $(x_0,y_0)$ along with $\mathbf{e}_0=(0,0.6,-0.8)$.
The top row shows the potential $V_s$ within the basin $V_s\leq 1/2$ (with boundaries shown as black solid and dashed lines corresponding to $e_z=-1$ and $e_z=+1$, respectively, as before) for different values of $C_s$ as determined from the stated initial conditions for $(x_0,y_0)$ along with $\mathbf{e}_0=(0,0.6,-0.8)$.
The bottom row illustrates the orbit corresponding to the stated initial conditions.
The left column illustrates an orbit whose basin is confined near the centre, the middle column illustrates an orbit whose basin is confined to an annular region, and the right column illustrates a basin which extends outside the cylindrical cross-section (red curve). 
The right most example is, of course, unphysical. 
%Recall that for cylindrical pipe flow there is an additional constant of motion which restricts the motion to that of a two dimensional dynamical system, so that the quasi-periodic orbits which are observed are confined more than the potential alone suggests.
The rotational symmetry leads to an additional constant of motion which restricts the motion to that of a two dimensional system, thus quasi-periodic, and results in tighter confinement than the potential alone suggests. % BH: needed to shorten this caption somewhat
}\label{fig:basins_spherical_cylinder}
\end{figure}

When $w(x,y)$ is the fluid velocity down a cylindrical pipe, e.g. in general $w(x,y)=W(1-(x^2+y^2)/R^2)$, then the rotational symmetry leads to an additional constant of motion which corresponds to the preservation of angular momentum around the $z$-axis.
Specifically, defining
\begin{equation*}
L_z:=x\dot{y}-y\dot{x}=xe_y-ye_x \,,
\end{equation*}
then for our model of active spherical particles it can be shown (in general) that
\begin{equation*}
\dot{L}_z = -\frac{e_z}{2}\left(x\frac{\partial w}{\partial y}-y\frac{\partial w}{\partial x}\right)\,.
\end{equation*}
For cylindrical pipe flow one has $\frac{\partial w}{\partial x}=-2Wx/R^2$ and $\frac{\partial w}{\partial y}=-2Wy/R^2$ from which it follows that $\dot{L}_z=0$.
Thus $L_z$ becomes an additional (independent) constant of motion and the dynamics reduce to that of a two dimensional system, which means there are a limited number of possible limit sets due to the Poincar\'e--Bendixson Theorem.
The dynamics of this special case were explored in detail by Z\"ottl and Stark \cite{ZottlStark2012}.
They illustrated that for certain initial conditions the motion is equivalent to that of a simple pendulum.
Our Hamiltonian formulation of the problem reveals a more general interpretation of the dynamics where there is a potential well $V_s$ possessing either a single minimum at the origin or else a circular locus of minima and that the orbits move up and down the potential well in a neighbourhood of the minima.
Several examples of basins and orbits are shown in \cref{fig:basins_spherical_cylinder}.

It can be reasoned that any $w(x,y)$ possessing rotational symmetry will result in $\dot{L}_z=0$, while in more general flows $L_z$ is not conserved.
As $L_z$ relates to the direction of motion around the origin, in cases of chaotic orbits the times at which $L_z=0$ indicate when changes in this direction occur and potentially provides a useful signature of each chaotic orbit.

\section{A Hamiltonian formulation of active prolate spheroidal particle motion}

In the case of a spheroidal particle with $G\ne0$, we must deal with all terms of the equations \cref{eqn:model_component_form}.
As in the case of a spherical particle, our goal is to first express the essential dynamics of the system via second order equations involving $x,y$ and then deriving a corresponding Hamiltonian.
To this end, first observe that
\begin{subequations}\label{eqn:general_second_order_equations}
\begin{align}
\ddot{x} &= \dot{e}_x = -\frac{1}{2}(1-G(1-2e_x^2))e_z\frac{\partial w}{\partial x}-Ge_xe_ye_z\frac{\partial w}{\partial y} \,, \\
\ddot{y} &= \dot{e}_y = -\frac{1}{2}(1-G(1-2e_y^2))e_z\frac{\partial w}{\partial y}-Ge_xe_ye_z\frac{\partial w}{\partial x} \,. \label{eqn:general_ddoty_equation}
\end{align}
\end{subequations}
In these expressions we may substitute $e_x=\dot{x}$ and $e_y=\dot{y}$, so provided we can again express $e_z$ in terms of $w(x,y)$ in some way (e.g. via a constant of motion similar to \cref{eqn:CoM_Geq0}) we can once again express the problem as a second order ODE.
We'll begin by examining the case of a prolate spheroid ($\gamma>1$ and thus $0<G<1$) and return with a more natural form for an oblate spheroidal particle later.

Examining \cref{eqn:dotez_general}, observe this can be equivalently expressed as
\begin{equation}\label{eqn:CoM_general_derivation}
\frac{\dot{e}_z}{1+G-2Ge_z^2}=\frac{1}{2}\frac{d}{dt}w(x,y)\,,
\end{equation}
and by integrating one finds that
\begin{equation*}
\frac{\tanh^{-1}(f_p(G)e_z)}{(1+G)f_p(G)}=\frac{1}{2}w(x,y)+\text{constant}
\end{equation*}
where $f_p(G)=\sqrt{2G/(1+G)}$.
We define the constant of integration via the initial conditions. 
Specifically, we choose to define
\begin{equation*}
C_p:=\tanh^{-1}(f_p(G)e_z(0))-\frac{1+G}{2}f_p(G)w(x(0),y(0))\,,
\end{equation*}
such that we obtain a second constant of motion (i.e. in addition to \cref{eqn:unit_e}) given by
\begin{equation}\label{eqn:CoM_general}
\tanh^{-1}(f_p(G)e_z(t))-\frac{1+G}{2}f_p(G)w(x(t),y(t))=C_p\,,
\end{equation}
for all $t\ge0$.

Using this additional constant of motion, we can eliminate $e_z$ from the second order equations \cref{eqn:general_second_order_equations} via the substitution
\begin{equation}\label{eqn:ez_sub}
e_z=\frac{1}{f_p(G)}\tanh\left(C_p+\frac{1+G}{2}f_p(G)w(x,y)\right)\,.
\end{equation}
That is, we now have
\begin{subequations}\label{eqn:second_order_prolate_model}
\begin{align}
\ddot{x} &= -\frac{1}{2f_p(G)}\left(\frac{\partial w}{\partial x}(1-G)+2G\dot{x}\left(\dot{x}\frac{\partial w}{\partial x}+\dot{y}\frac{\partial w}{\partial y}\right)\right)\tanh\left(C_p+\frac{1+G}{2}f_p(G)w\right) \,, \\
\ddot{y} &= -\frac{1}{2f_p(G)}\left(\frac{\partial w}{\partial y}(1-G)+2G\dot{y}\left(\dot{x}\frac{\partial w}{\partial x}+\dot{y}\frac{\partial w}{\partial y}\right)\right)\tanh\left(C_p+\frac{1+G}{2}f_p(G)w\right) \,.
\end{align}
\end{subequations}

We now seek a Hamiltonian/Lagrangian formulation of this system.
Let $P_x,P_y$ be (generalised) momenta for which
$$\mathcal{H}(x,y,P_x,P_y)+\mathcal{L}(x,y,\dot{x},\dot{y})=\dot{x}P_x+\dot{y}P_y\,,$$
with $\mathcal{H}$ denoting the Hamiltonian and $\mathcal{L}$ denoting the Lagrangian.
We suppose that the Hamiltonian has the form
$$\mathcal{H}=\frac{P_x^2+P_y^2}{2F_p(w(x,y))}+V_p(w(x,y))\,,$$
with functions $F_p(w),V_p(w)$ to be determined.
Subsequently, Hamilton's equations give
$$\dot{x}=\frac{\partial \mathcal{H}}{\partial P_x}=\frac{P_x}{F_p(w(x,y))} \quad\Rightarrow\quad P_x=\dot{x}F_p(w(x,y))\,,$$
and similarly 
$$\dot{y}=\frac{\partial \mathcal{H}}{\partial P_y}=\frac{P_y}{F_p(w(x,y))} \quad\Rightarrow\quad P_y=\dot{y}F_p(w(x,y))\,.$$
The corresponding Lagrangian then has the form
$$\mathcal{L}=\frac{\dot{x}^2+\dot{y}^2}{2}F_p(w(x,y))-V_p(w(x,y))\,,$$
from which we consistently find
$$P_x:=\frac{\partial \mathcal{L}}{\partial \dot{x}}=\dot{x}F_p(w(x,y))\,,$$
and similarly for $P_y$.

Now, we claim that 
\begin{equation*}
F_p(w)=\cosh\bigg(C_p+\frac{1+G}{2}f_p(G)w\bigg)
\end{equation*}
is a suitable $F_p$ for formulating a Hamiltonian.
This comes from first examining
\begin{align*}
\dot{P}_x
&=\frac{d}{dt}[\dot{x}F_p(w(x,y))] \\
&=\ddot{x}F_p(w(x,y))+\dot{x}F_p^\prime(w(x,y))(\dot{x}w_x(x,y)+\dot{y}w_y(x,y)) \\
&=\bigg(-\frac{1}{2}w_x\big[1-G+2G\dot{x}(\dot{x}w_x+\dot{y}w_y)\big]e_z\bigg)F_p(w)+\dot{x}F_p^\prime(w)(\dot{x}w_x+\dot{y}w_y)\\
&=-\frac{1}{2}w_x(1-G)e_zF_p(w)+(F_p^\prime(w)-Ge_zF_p(w))\dot{x}(\dot{x}w_x+\dot{y}w_y) \,.
\end{align*}
Here we've left $e_z$ without substituting \cref{eqn:ez_sub} to simplify the expression, and also utilised the shorthand $w_x:=\frac{\partial w}{\partial x}$ and $w_y:=\frac{\partial w}{\partial y}$.
Observe that if we had $F_p^\prime(w)-Ge_zF_p(w)=0$ it would eliminate the terms involving $\dot{x},\dot{y}$ factors and the remaining term could be expressed as $-\frac{1-G}{2G}w_x F_p^\prime(w)$ which is proportional to $\frac{\partial}{\partial x}F_p(w(x,y))$.
Ultimately, we also wish to identify the above equation for $\dot{P}_x$ with the corresponding Hamilton equation 
\begin{align*}
\dot{P}_x=-\frac{\partial \mathcal{H}}{\partial x} 
&=\frac{P_x^2+P_y^2}{2F_p(w(x,y))^2}F_p^\prime(w(x,y))w_x(x,y)-V_p^\prime(w(x,y))w_x \\
&=\frac{e_x^2+e_y^2}{2}F_p^\prime(w)w_x-V'(w)w_x \\
&=\frac{1-e_z^2}{2}F_p^\prime(w)w_x-V_p^\prime(w)w_x \,.
\end{align*}
Doing so will enable us to find $V_p(w)$.
An examination of $\dot{P}_y$ leads to the same conclusions.

Thus, let us now choose $F_p(w)$ to be such that $F_p^\prime(w)=Ge_zF_p(w)$.
This leads to
\begin{equation*}
\frac{dF_p}{F_p}=\frac{G}{f_p(G)}\tanh\bigg(C_p+\frac{1+G}{2}f_p(G)w\bigg)\,dw \quad\Rightarrow\quad F_p(w)=A\cosh\bigg(C_p+\frac{1+G}{2}f_p(G)w\bigg) \,,
\end{equation*}
and without loss of generality we choose the integration factor to be $A=1$.

Equating the two expressions for $\dot{P}_x$ now leads to
$$-\frac{1-G}{2G}w_xF_p^\prime(w)=\dot{P}_x=\frac{1-e_z^2}{2}F_p^\prime(w)w_x-V_p^\prime(w)w_x$$
which, on noting that $w_x$ is generally non-zero, can be re-arranged to 
\begin{align*}
V_p^\prime(w) &=\left(\frac{1-e_z^2}{2}+\frac{1-G}{2G}\right)F_p^\prime(w) \\
&=\frac{1-Ge_z^2}{2G}F_p'(w) \\
&=\frac{1}{2G}F_p^\prime(w)\left[1-\frac{G}{f_p(G)^2}\tanh\left(C_p+\frac{1+G}{2}f_p(G)w(x,y)\right)^2\right] \\
&=\frac{1}{2G}F_p^\prime(w)\left[1-\frac{1+G}{2}\frac{F_p(w)^2-1}{F_p(w)^2}\right] \\
&=\frac{1}{2G}F_p^\prime(w)\left[\frac{1-G}{2}+\frac{1+G}{2F_p(w)^2}\right] \,.
\end{align*}
By integrating with respect to $w$ we can deduce that an appropriate function for $V_p(w)$ is given by
\begin{equation*}
V_p(w) = \frac{1}{2G}\left[\frac{1-G}{2}F_p(w)-\frac{1+G}{2F_p(w)}\right]+\frac{1}{2}\,.
\end{equation*}
Note that the addition of $1/2$ at the end isn't strictly necessary, but has been done such that the relevant range of $V_p$ is identical to the relevant range of $V_s$ (as will be seen later).

To summarise, an appropriate Hamiltonian for our dynamical system is
\begin{equation}
\mathcal{H}=\frac{P_x^2+P_y^2}{2F_p(w)}+\frac{1}{2G}\left[\frac{1-G}{2}F_p(w)-\frac{1+G}{2F_p(w)}\right]+\frac{1}{2}\,.
\end{equation}
We can double check this is correct by showing that applying Hamilton's equations re-produces \cref{eqn:general_ddoty_equation} and that $\mathcal{H}$ conserved throughout the motion, see \cref{app:verification}.
It is important to note that this Hamiltonian is not independent from the existing constants of motion. 
Specifically, it can be shown that $2\mathcal{H}-1=(\|\mathbf{e}\|^2-1)F_p(w)$ (thus every orbit maintains $\mathcal{H}=1/2$).

\begin{figure}
\begin{center}
\includegraphics{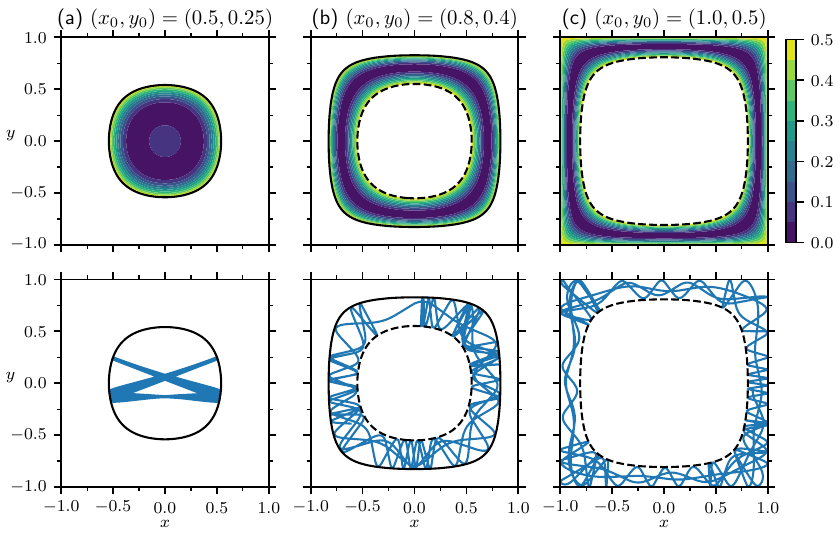}
\end{center}
\caption{Several examples of potentials and orbits of an active prolate spheroidal particle with $\gamma=2$ ($G=0.6$) in square duct flow, specifically utilising \cref{eqn:square_poiseuille}. 
The top row shows the potential $V_p$ within the basin $V_p\leq 1/2$ for different values of $C_p$ as determined from the stated initial conditions for $(x_0,y_0)$ along with $\mathbf{e}_0=(0,0,-1)$.
Black solid and dashed lines show where $V_p=1/2$ corresponding to particle motion perpendicular to the cross-section, specifically with $e_z=-1$ and $e_z=+1$, respectively.
The bottom row illustrates the orbit corresponding to the stated initial conditions.
The left column illustrates an orbit whose basin is confined near the centre, the middle column illustrates an orbit whose basin is confined to an annular region, and the right column illustrates a basin which is confined to a neighbourhood of the duct wall.
Observe that the confinement is tight.
The initial conditions in each column are identical to those used for a spherical particle in \cref{fig:basins_spherical_square}.
}\label{fig:basins_prolate_square}
\end{figure}

Our formulation as a system of second order equations, with a corresponding Hamiltonian, has all of the advantages described above in the case of an active spherical particle.
However, we remark that this particular Hamiltonian is non-separable, so some care must be taken with the use of symplectic integrators (many of which assume a separable Hamiltonian).
Additionally, the constant of motion \cref{eqn:CoM_general} may be used to determine a region of the duct to which an orbit will be confined.
Specifically, given a value of $C_p$ from the initial conditions, then as $|e_z|\leq1$ it follows that
\begin{equation*}
2\frac{-\tanh^{-1}(f_p(G))-C_p}{(1+G)f_p(G)}\leq w(x(t),y(t))\leq 2\frac{\tanh^{-1}(f_p(G))-C_p}{(1+G)f_p(G)} \,,
\end{equation*}
or equivalently
\begin{multline*}
w(x(0),y(0))-2\frac{\tanh^{-1}(f_p(G)e_z(0))+\tanh^{-1}(f_p(G))}{(1+G)f_p(G)}\leq w(x(t),y(t)) \\
\leq w(x(0),y(0))-2\frac{\tanh^{-1}(f_p(G)e_z(0))-\tanh^{-1}(f_p(G))}{(1+G)f_p(G)} \,.
\end{multline*}
Thus the motion of the particle is confined between these two level sets of $w(x,y)$.
Moreover, it can be shown that the potential $V_p(w)$ remains within the interval $[0,1/2]$, see \cref{app:potential_properties}, which provides an alternative way to examine the confinement of the orbit of a prolate spheroidal active particle analogous to that of a spherical active particle.
Several examples of the orbit of an active prolate spheroidal particle within a square duct, alongside their basin, is shown in \cref{fig:basins_prolate_square}.

\subsection{General classification of stationary points for prolate spheroidal active particles}

We now conduct a general classification of stationary points of \cref{eqn:general_second_order_equations}, assuming $0<G<1$.
Once again, their location and properties can generally be inferred from the potential $V_p$, but given the interplay between the initial orientation of the particle and the specific nature of $V_p$, we find it more straightforward to examine \cref{eqn:general_second_order_equations} as a first order system.

Let $\mathbf{v}=[x,y,P_x,P_y]$ and consider the first order system describing $\dot{\mathbf{v}}$ utilising \cref{eqn:general_second_order_equations}.
The conditions $\dot{x}=\dot{y}=0$ imply that $e_x=e_y=0$, and consequently $e_z=\pm1$.
The condition $\dot{P}_x=0$ can then be reduced to $-V_p^\prime(w)w_x=0$ from which it can then be inferred that either $w_x=0$ or $F_p^\prime(w)=0$.
The latter can only occur if $e_z=0$ which contradicts $e_z=\pm1$, thus $w_x=0$ is the only possibility.
By similar reasoning $\dot{P}_y=0$ implies $w_y=0$.
Thus, once again, stationary points of \cref{eqn:general_second_order_equations} occur at critical points of $w(x,y)$ with particle orientations of $\mathbf{e}=(0,0,\pm1)$.

To examine the stability of the stationary points we can form the Jacobian and analyse its eigenvalues.
For the prolate spheroidal particle, at the stationary points the Jacobian with respect to $\mathbf{v}$ reduces to
$$
\begin{bmatrix}0&0&1/F_p(w)&0\\0&0&0&1/F_p(w)\\a&b&0&0\\b&c&0&0\end{bmatrix}\,, 
\quad\text{where}\quad
\begin{array}{l} 
a = -\frac{1-G}{2}e_z F_p(w)w_{xx} \,,\\[2pt]
b = -\frac{1-G}{2}e_z F_p(w)w_{xy} \,,\\[2pt]
c = -\frac{1-G}{2}e_z F_p(w)w_{yy} \,,
\end{array}
$$
noting we've utilised $e_z$ in these expressions as a shorthand for \cref{eqn:ez_sub}.
The eigenvalues then satisfy
$$\lambda^2=\frac{a+c}{2F_p(w)}\pm\frac{1}{2F_p(w)}\sqrt{(a+c)^2-4(ac-b^2)} \,.$$
On substitution of $a,b,c$ we find that all factors of $F_p(w)$ cancel and the $\lambda^2$ are identical to those in the case of a spherical particle apart from multiplication by a factor $1-G$.
Thus, the eigenvalues are $\sqrt{1-G}$ times those of the spherical particle case and identical conclusions about the stability of stationary points can be made.
To summarise, elongation of the particle along its swimming axis does not change the nature of the stationary points.

\subsection{Special case of an active prolate spheroidal particle in cylindrical pipe flow}

\begin{figure}
\begin{center}
\includegraphics{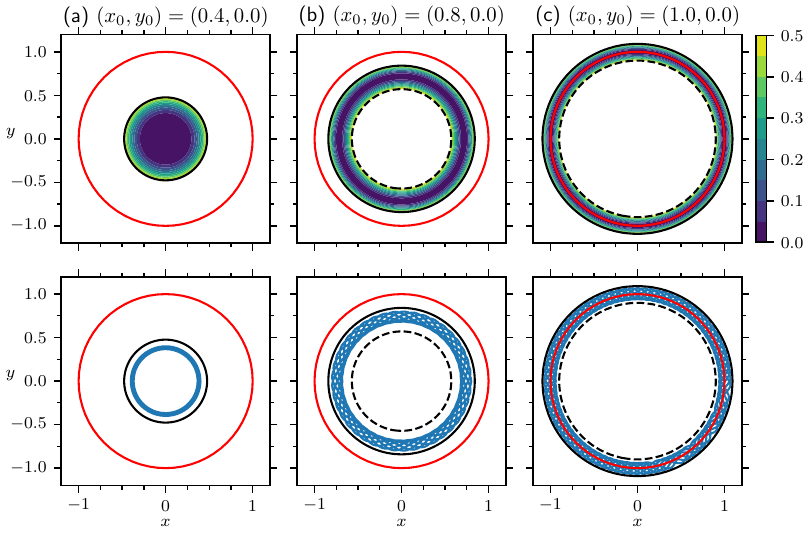}
\end{center}
\caption{Several examples of potentials and orbits of an active prolate spheroidal particle with $\gamma=2$ ($G=0.6$) in cylindrical pipe flow, specifically utilising \cref{eqn:circle_poiseuille} with $W=10$. 
%The top row shows the boundaries of the basin (black solid and dashed lines corresponding to $e_z=-1$ and $e_z=+1$, respectively, as before) and the potential $V_p$ within the basin for different values of $C_p$ as determined from the stated initial conditions for $(x_0,y_0)$ along with $\mathbf{e}_0=(0,0.6,-0.8)$.
The top row shows the potential $V_p$ within the basin $V_p\leq 1/2$ (with boundaries shown as black solid and dashed lines corresponding to $e_z=-1$ and $e_z=+1$, respectively, as before) for different values of $C_p$ as determined from the stated initial conditions for $(x_0,y_0)$ along with $\mathbf{e}_0=(0,0.6,-0.8)$.
The bottom row illustrates the orbit corresponding to the stated initial conditions.
The left column illustrates an orbit whose basin is confined near the centre, the middle column illustrates an orbit whose basin is confined to an annular region, and the right column illustrates a basin which is confined to a neighbourhood of the duct wall.
The right most example is, of course, unphysical.
Observe that the confinement of orbits is tighter than is predicted by $V_p$ due to the existence of an additional constant of motion (namely $L_z$) which further constrains the dynamics.
The initial conditions in each column are identical to those used for a spherical particle in \cref{fig:basins_spherical_cylinder} for comparison.}\label{fig:basins_prolate_cylinder}
\end{figure}

When $w(x,y)$ is the fluid velocity down a cylindrical pipe, e.g. in general $w(x,y)=W(1-(x^2+y^2)/R^2)$, then there is an additional constant of motion which comes about due to the preservation of angular momentum around the $z$-axis.
Specifically, for the prolate spheroidal particle we define
$$L_z:=xP_y-yP_x=(xe_y-ye_x)F_p(w) \,.$$
Then for a general flow it can be shown that
\begin{align*}
\dot{L}_z &= (\dot{x}e_y+x\dot{e}_y-\dot{y}e_x-y\dot{e}_x)F_p(w)+(xe_y-ye_x)F_p^\prime(w)(\dot{x}w_x+\dot{y}w_y) \\
&= (x\dot{e}_y-y\dot{e}_x)F_p(w)+(xe_y-ye_x)Ge_zF_p(w)(\dot{x}w_x+\dot{y}w_y) \,,
\end{align*}
noting we've utilised the relation $F_p^\prime(w)=Ge_zF_p(w)$.
Substituting $\dot{e}_x$ and $\dot{e}_y$ then yields
\begin{align*}
\dot{L}_z &= \bigg(-\frac{x}{2}w_y(1-G)-Gxe_y(e_xw_x+e_yw_y) 
\\&\qquad +\frac{y}{2}w_x(1-G)+Gye_x(e_xw_x+e_yw_y) 
\\&\qquad+G(xe_y-ye_x)(e_xw_x+e_yw_y)\bigg)e_zF_p(w) \\
&= \frac{1-G}{2}(yw_x-xw_y)e_zF_p(w) \,.
\end{align*}
For a cylindrical pipe one has $w_x=-2Wx/R^2$ and $w_y=-2Wy/R^2$, from which it follows that $\dot{L}_z=0$.
Moreover, it can be reasoned that any $w(x,y)$ possessing rotational symmetry will result in $\dot{L}_z=0$.
Thus, once more, $L_z$ becomes an additional (i.e. independent) constant of motion and the dynamics of the system reduces to that of a continuous two dimensional dynamical system, which means there are a limited number of possible limit sets due to the Poincare--Bendixson Theorem.
In other words, the non-spherical shape of the particle does not significantly alter the complexity of the dynamics in this special case.
Several examples of the orbit of an active prolate spheroidal particle within a square duct, alongside their basin, is shown in \cref{fig:basins_prolate_cylinder}.

\section{A Hamiltonian formulation of active oblate spheroidal particle motion}\label{sec:oblate_spheroids}

For the case of an active oblate spheroidal particle ($0<\gamma<1$ and $-1<G<0$) we can take an identical approach to that for a prolate spheroidal particle. 
Observe that in formulating the constant of motion \cref{eqn:CoM_general} we find that with $-1<G<0$ then $f_p(G)$ is imaginary and it is more natural to use (standard) trigonometric functions in the description rather than the hyperbolic ones. 
Specifically, the first point of difference which occurs is that the solution of \cref{eqn:CoM_general_derivation} is more naturally described as
$$\frac{\tan^{-1}(f_o(G)e_z)}{(1+G)f_o(G)}=\frac{1}{2}w(x,y)+\text{constant}\,,$$
where we define $f_o(G)=\sqrt{-2G/(1+G)}$ to avoid the need to consider an imaginary valued $f_p(G)$.
Subsequently, we define
$$C_o=\tan^{-1}(f_o(G)e_z(0))-\frac{1+G}{2}f_o(G)w(x(0),y(0))\,,$$
which leads to the general substitution
\begin{equation}
e_z=\frac{1}{f_o(G)}\tan\left(C_0+\frac{1+G}{2}f_o(G)w(x,y)\right)\,. \label{eqn:ez_oblate}
\end{equation}
Using this substitution allows us to express the dynamics as a coupled second order system
\begin{subequations}\label{eqn:second_order_oblate_model}
\begin{align}
\ddot{x} &= -\frac{1}{2f_o(G)}\left(\frac{\partial w}{\partial x}(1-G)+2G\dot{x}\left(\dot{x}\frac{\partial w}{\partial x}+\dot{y}\frac{\partial w}{\partial y}\right)\right)\tan\left(C_o+\frac{1+G}{2}f_o(G)w\right) \,, \\
\ddot{y} &= -\frac{1}{2f_o(G)}\left(\frac{\partial w}{\partial y}(1-G)+2G\dot{y}\left(\dot{x}\frac{\partial w}{\partial x}+\dot{y}\frac{\partial w}{\partial y}\right)\right)\tan\left(C_o+\frac{1+G}{2}f_o(G)w\right) \,.
\end{align}
\end{subequations}
Continuing on, we again propose the ansatz $\mathcal{H}=\frac{P_x^2+P_y^2}{2F_o(w(x,y))}+V_o(w(x,y))$ from which we define momenta $P_x=\dot{x}F_o(w)$ and $P_y=\dot{y}F_o(w)$. 
Through an examination of $\dot{P}_x$ we similarly choose $F_o$ such that $F_o^\prime(w)=Ge_zF_o(w)$.
With $e_z$ as in \cref{eqn:ez_oblate} this leads to
$$F_o(w)=\cos\left(C_o+\frac{1+G}{2}f_o(G)w(x,y)\right)\,.$$
Finally, one puts together the two distinct expressions for $\dot{P}_x$ to deduce that
$$V_o(w)=\frac{1}{2G}\left[\frac{1-G}{2}F_o(w)-\frac{1+G}{2F_o(w)}\right]+\frac{1}{2}$$
is a suitable potential function.
Observe that this $V_o(w)$ has the same form as $V_p(w)$ but utilises $F_o(w)$ in place of $F_p(w)$.
Although it may initially appear that the standard trigonometric functions like $\tan$ will introduce issues involving singularities, the fact that the particle remains confined to the region with $V_o\in[0,1/2]$ keeps the arguments of these functions away from such singularities.

\begin{figure}
\begin{center}
\includegraphics{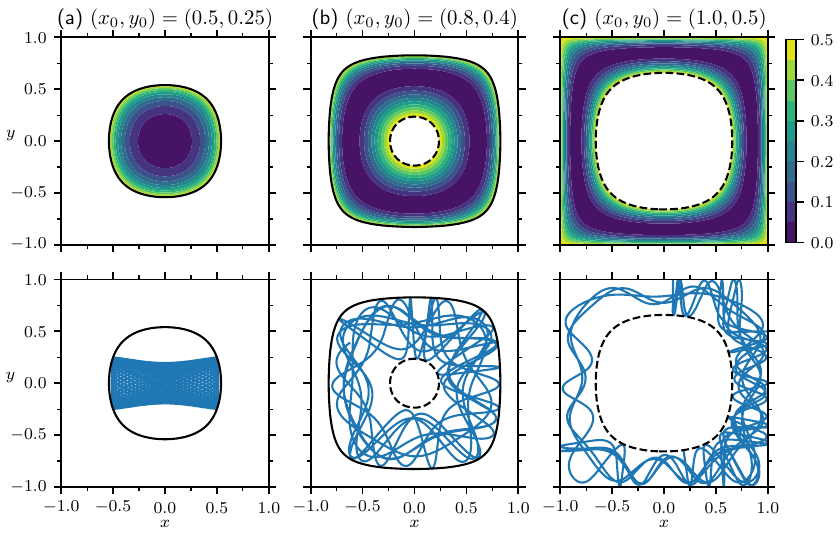}
\end{center}
\caption{Several examples of potentials and orbits of an active prolate spheroidal particle with $\gamma=0.5$ ($G=-0.6$) in square duct flow, specifically utilising \cref{eqn:square_poiseuille}. 
The top row shows the potential $V_o$ within the basin $V_o\leq 1/2$ for different values of $C_o$ as determined from the stated initial conditions for $(x_0,y_0)$ along with $\mathbf{e}_0=(0,0,-1)$.
Black solid and dashed lines show where $V_o=1/2$ corresponding to particle motion perpendicular to the cross-section, specifically with $e_z=-1$ and $e_z=+1$, respectively.
The bottom row illustrates the orbit corresponding to the stated initial conditions.
The left column illustrates an orbit whose basin is confined near the centre, the middle column illustrates an orbit whose basin is confined to an annular region, and the right column illustrates a basin which is confined to a neighbourhood of the duct wall.
Observe that the confinement is tight.
The initial conditions in each column are identical to those used for a spherical particle in \cref{fig:basins_spherical_square} and a prolate spheroidal particle in \cref{fig:basins_prolate_square}.
}\label{fig:basins_oblate_square}
\end{figure}

Of course, these modifications in the case of an oblate spheroidal particle are also readily obtained via the relationship between the hyperbolic and standard trigonometric functions given imaginary arguments.
As such, the general observations about modelling active prolate spheroidal particle motion via its Hamiltonian formulation can be applied to the oblate case with the appropriate modifications.
For example, the specific case of a flow $w(x,y)$ possessing rotational symmetry leads to the preservation of $L_z=(xe_y-ye_x)F_o(w)$.
Additionally, the general classification of stationary points is identical.
Several examples of the orbit of an active oblate spheroidal particle within a square duct, alongside their basin, is shown in \cref{fig:basins_oblate_square}.

\section{Comparison between spherical and spheroidal active particles}\label{sec:comparisons}

\begin{figure}
\begin{subfigure}{\textwidth}
\centering
\includegraphics{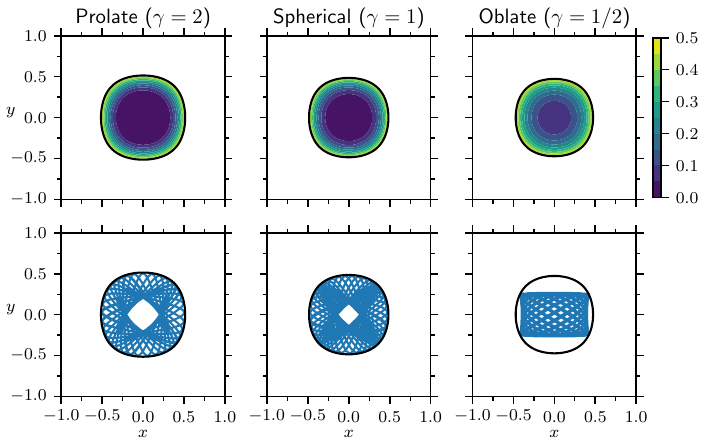}
\caption{Potentials and orbits with initial conditions $(x_0,y_0)=(0.4,0.2)$ and $\mathbf{e}_0=(0.36,0.48,-0.8)$.}\label{fig:basin_comparisons_a}
\end{subfigure}
\begin{subfigure}{\textwidth}
\centering
\includegraphics{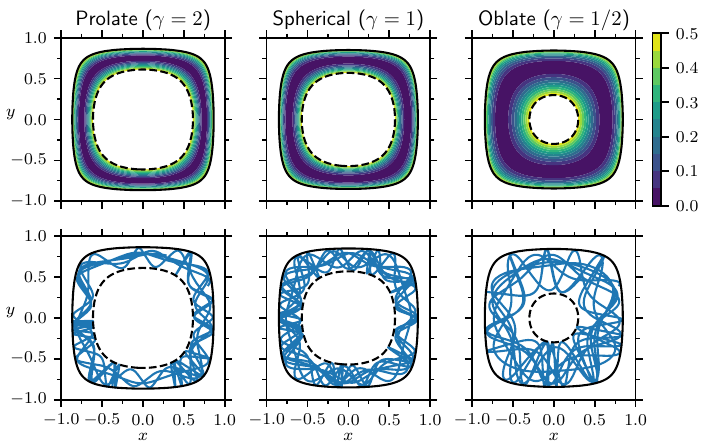}
\caption{Potentials and orbits with initial conditions $(x_0,y_0)=(0.8,0.4)$ and $\mathbf{e}_0=(0.36,0.48,-0.8)$.}\label{fig:basin_comparisons_b}
\end{subfigure}\caption{
Side by side comparison of the potential and corresponding orbit across three different examples of spheroidal particle shapes ($\gamma=2,1,1/2$) for two distinct initial conditions.}\label{fig:basin_comparisons}
\end{figure}

Here we briefly illustrate some of the differences between basins and orbits of active particles produced by the model in the three different cases, i.e. spherical ($\gamma=1$), prolate spheroidal ($\gamma>1$) and oblate spheroidal ($0<\gamma<1$).
\cref{fig:basin_comparisons} compares the basins and orbit for two different initial conditions across the three cases. 
Some general observations to be made from results across many more initial conditions than here shown are: (a) for basins not containing the origin, oblate particles generally have a larger basin than spherical particles which in turn have a slightly larger basin than prolate particles (see \cref{fig:basin_comparisons_b}), (b) prolate particles typically oscillate at a lower frequency than both spherical and oblate particles (see \cref{fig:oscillation_comparison}), and (c) there are a range of initial conditions where the orbit of the oblate particle appears to be much more regular than that of the spherical and prolate particles (see \cref{fig:component_comparison2,fig:component_comparison3,fig:component_comparison4}).

\begin{figure}
\centering
\includegraphics[]{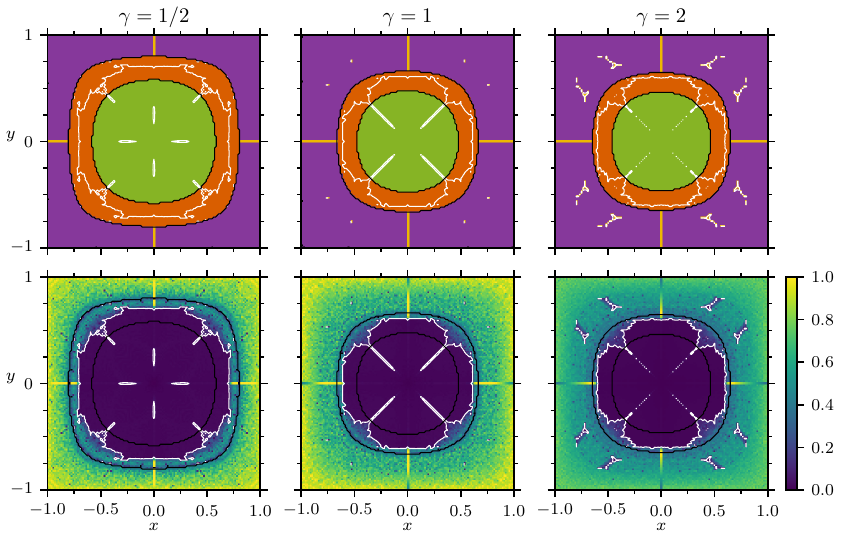}
\caption{Classification of active particle trajectories (top row) and estimate of the largest Lyapunov exponent (LLE) (bottom row) in the initial position $(x(0),y(0))$ plane for $\gamma=1/2$ (left column), $\gamma=1$ (middle column) and $\gamma=2$ (right column). 
Other parameters are fixed to $W=10$ and $\mathbf{e}(0)=(0,0,-1)$.
Orbits are classified in the top row according to the discussion in \cref{sec:spherical_hamiltonian} as follows.
Green \textcolor{mygreen}{\Large \textbullet}: simple swinging/wandering with trajectories for which $e_z\leq 0$ for all $t$. 
Red \textcolor{myred}{\Large \textbullet}: complex swinging/wandering with trajectories for which $e_z\leq E$ for all $t$ for some $0<E<1$. 
Purple \textcolor{mypurple}{\Large \textbullet}: tumbling with trajectories for which $e_z=1$ can be achieved. 
Yellow \textcolor{myyellow}{\Large \textbullet}: tumbling with trajectories which are additionally confined in a rectangular region of the cross section which may cross either centreline ($x=0$, $y=0$), or neither, but not both. 
Black lines separating simple swinging/wandering, complex swinging/wandering and tumbling are shown in both plots for each $\gamma$.
Similarly, white contours showing where the LLE equals $0.05$ are shown in both plots for each $\gamma$.
LLE values greater than $1$ were achieved but have been clipped to highlight detail at smaller values.
}\label{fig:ic_compare}
\end{figure}

In \cite{ValaniHardingStokes2024} we carried out a classification of orbits of active spherical particles suspended in flow through a square duct, specifically utilising the approximation $w(x,y)=W(1-x^2)(1-y^2)$ for the fluid velocity, with $W=10$.
The classification distinguished between 6 different types of orbits: central swinging (confined to $|x|,|y|<1/4$), horizontal or vertical swinging (confined with $|y|<1/4$ or $|x|<1/4$, respectively), off-centred trapping (confined to quadrant of the duct, e.g. $x,y>0$), tumbling (orbits that surround but do not contain the origin), and wandering (orbits which appear unrestricted around a sufficiently large neighbourhood of the origin).
In \cref{sec:spherical_hamiltonian} we introduced a refined definition of tumbling, swinging and wandering motions for which classification is made more robust via the potential which arises from our Hamiltonian formulation.
Although discussed in the context of the Hamiltonian for a spherical active particle, it can be readily applied to the prolate and oblate spheroidal active particles as well.

In \cref{fig:ic_compare} we apply the new classification to the orbits observed for spheroidal particles with a fixed initial orientation $\mathbf{e}(0)=(0,0,-1)$ and varying initial position in the cross-section $(x(0),y(0))$ across each of the three cases $\gamma=1/2,1,2$ corresponding to oblate, spherical and prolate shaped particles, respectively.
As in \cite{ValaniHardingStokes2024}, we further distinguish between a sub-class of tumbling motion, off-centre trapping, in which the particle trajectory may cross at either of the axes ($x=0$ or $y=0$), or neither, but not both.
On the other hand, while the classification of \cite{ValaniHardingStokes2024} also distinguished between central, vertical and horizontal swinging motions, we have not done so here.
Further, we have used the series solution for Poiseuille flow, that is \cref{eqn:square_poiseuille}, truncated beyond $n=10$.
Each classification plot has been paired with a plot estimating the largest Lyapunov exponent (LLE) for each initial condition. 
A white contour where the LLE is equal to $0.05$ is provided as a useful threshold to help distinguish between chaotic and periodic or quasi-periodic behaviour.

There are some broad qualitative similarities across the three cases, but also some distinct differences which we highlight here.
Starting with some commonalities, regions of simple swinging/wandering (green) generally have quite small LLE indicating that the trajectories of particles with an initial position in this region are typically periodic or quasi-periodic.
As such, simple swinging/wandering might be more simply described as simple swinging.
In regions of complex swinging/wandering, we generally observe an increase in the LLE indicating that the trajectories of particles with an initial position in this region are typically chaotic.
Tumbling trajectories generally have a larger LLE indicating they are mostly chaotic excepting some of those which are classified as featuring off-centre trapping\footnote{We note that the result for $\gamma=1$ does not include the erroneous regions of off-centre trapping that were observed in \cite{ValaniHardingStokes2024} as a result of a coding error. However, regions of off-centre trapping are still realised for $\gamma=1$ with larger values of $W$.}.
The trajectories along the axes of symmetry which have been classified as off-centre trapping typically have quite a large LLE owing to a large expansion coefficient orthogonal to the line on which the trajectory sits.
(It should be noted that our calculations don't capture orbits which should remain on the diagonal lines of symmetry because the fluid velocity \cref{eqn:square_poiseuille} does not possess such symmetry when truncated.)

Now we briefly discuss some of the differences between the three cases.
The prolate particle ($\gamma=2$) shows a classification and LLE structure which has more qualitative similarities to that of the spherical particle, albeit with the slightly enlarged region of tumbling motion and also some regions of off-centre trapping orbits away from the $x$- and $y$-axes.
Observe the horizontally and vertically aligned structure in the region of smallest LLE, which might be sub-classified into vertical and horizontal swinging (and central swinging where they overlap). 
The oblate particle ($\gamma=1/2$) has a larger region of swinging/wandering orbits, owing to the larger basin which encompasses the origin more often, largely at the expense of the regions of tumbling orbits.
The central region where the LLE is small is aligned with the diagonals in the oblate case, unlike that in the spherical and prolate cases.
Our explorations generally suggest that oblate particles have an affinity for orbits in the neighbourhood of a diagonal. 
Thus, it might be appropriate introduce a sub-classification of diagonal swinging, i.e. in addition to the horizontal and vertical swinging sub-classifications used previously \cite{ValaniHardingStokes2024}.
It should be noted that the oblate and prolate particles potentially exhibit some more exotic orbits, such as that in \cref{fig:basins_prolate_square}(a), which may warrant a distinct classification. 
Significantly more exploration could be done, for example investigating changes in classification for alternative initial orientations and/or with changes in the maximum fluid velocity $W$, but we leave this for future work.

%\RV{Perhaps, highlight here or in conclusions that our Hamiltonian formulation of the system implies that overdamped active motion in a channel can be converted into inertial passive particle motion in a modified potential. Hence, in some sense, activity gets converted to inertia plus and external initial-condition dependent potential. Might also be good to cite some experimental papers where people have looked at active particles in microfluidic channels.} %%% BH: I'm not sure I really understand this comment, might be easier for Rahil to directly add some text here?

\begin{figure}
\centering
\includegraphics[]{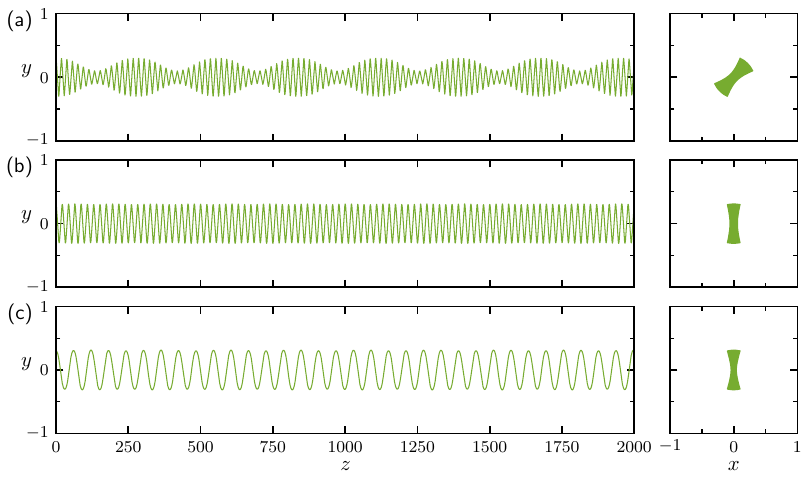}
\caption{Comparison of trajectories for active ellipsoidal particles with (a) $\gamma=1/5$, (b) $\gamma=1$ and (c) $\gamma=5$. 
Projections in two different planes are shown: (left) the $y-z$ plane and (right) the $y-x$ plane. 
The colour of the trajectory is based on the classification of \cref{fig:ic_compare}. 
In each case $W=10$ and the initial conditions are $\mathbf{x}(0)=(0.1,0.3,0)$ and $\mathbf{e}(0)=(0,0,-1)$. 
}\label{fig:oscillation_comparison}
\end{figure}

\begin{figure}
\centering
\includegraphics[]{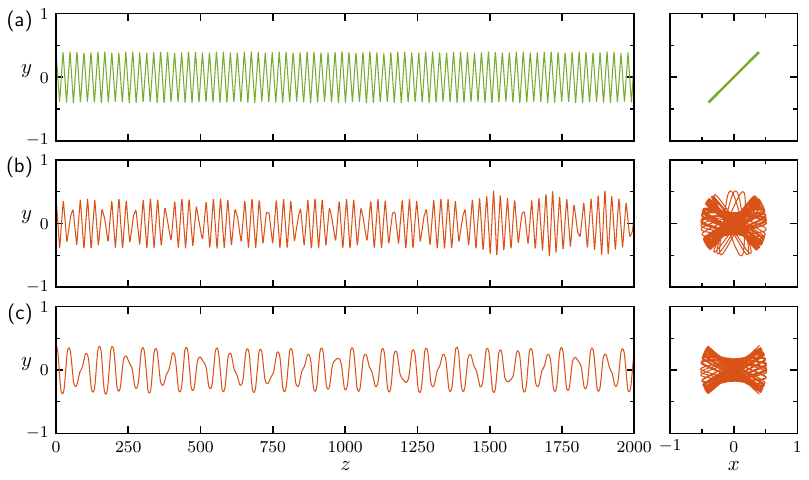}
\caption{Comparison of trajectories for active ellipsoidal particles with (a) $\gamma=1/5$, (b) $\gamma=1$ and (c) $\gamma=5$. 
Projections in two different planes are shown: (left) the $y-z$ plane and (right) the $y-x$ plane. 
The colour of the trajectory is based on the classification of \cref{fig:ic_compare}.
In each case $W=10$ and the initial conditions are $\mathbf{x}(0)=(0.4,0.38,0)$ and $\mathbf{e}(0)=(0,0,-1)$. 
}\label{fig:component_comparison2}
\end{figure}

\begin{figure}
\centering
\includegraphics[]{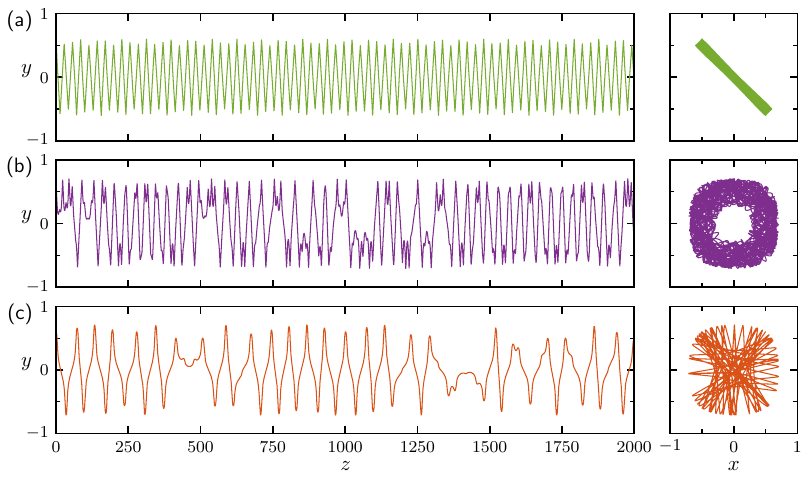}
\caption{Comparison of trajectories for active ellipsoidal particles with (a) $\gamma=1/5$, (b) $\gamma=1$ and (c) $\gamma=5$. 
Projections in two different planes are shown: (left) the $y-z$ plane and (right) the $y-x$ plane. 
The colour of the trajectory is based on the classification of \cref{fig:ic_compare}. 
In each case $W=10$ and the initial conditions are $\mathbf{x}(0)=(-0.5,0.6,0)$ and $\mathbf{e}(0)=(0,0,-1)$. 
}\label{fig:component_comparison3}
\end{figure}

\begin{figure}
\centering
\includegraphics[]{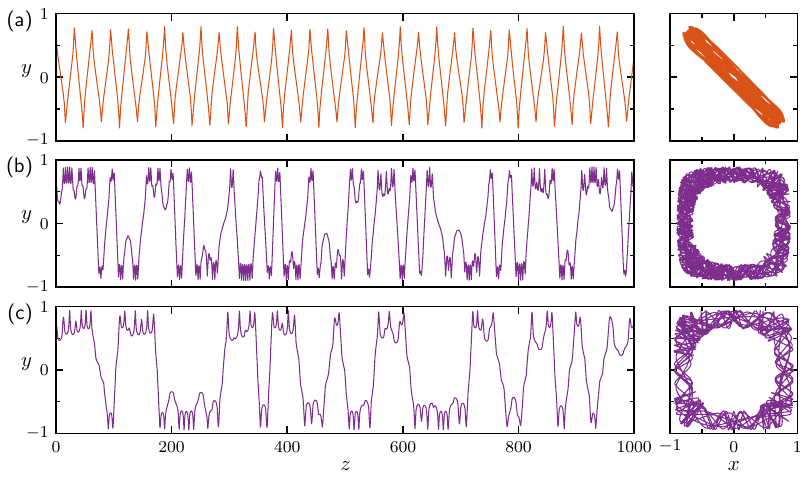}
\caption{Comparison of trajectories for active ellipsoidal particles with (a) $\gamma=1/5$, (b) $\gamma=1$ and (c) $\gamma=5$. 
Projections in two different planes are shown: (left) the $y-z$ plane and (right) the $y-x$ plane. 
The colour of the trajectory is based on the classification of \cref{fig:ic_compare}. 
In each case $W=10$ and the initial conditions are $\mathbf{x}(0)=(-0.75,0.7,0)$ and $\mathbf{e}(0)=(-1/\sqrt{2},0,-1/\sqrt{2})$. 
}\label{fig:component_comparison4}
\end{figure}

\section{Conclusions}

The main contribution of this work has been to show that the active particle model of Z\"ottl and Stark, in the case of both spherical \cite{ZottlStark2012} and prolate spheroidal \cite{ZottlStark2013} particles, can be formulated as a Hamiltonian system for a general flow $\mathbf{u}_f=w(x,y)\mathbf{k}$.
An interesting feature of this system is that the nature of the potential function depends on the initial conditions of the system.
We have briefly outlined some of the advantages of this formulation, both in its utility and the insights gained into the dynamics of motion. 
In particular, this formulation circumvents issues that potentially arise from parametrising the direction vector $\mathbf{e}$ in spherical coordinates, can be solved efficiently with integrators tailored to second order problems (and potentially using symplectic integrators to preserve constants of motion over longer integration times), and the derived potential provides insights into the confinement of orbits and allows for a more robust classification of trajectories.
The mapping of the active particle-fluid system onto a Hamiltonian system shows that an active self-propelled particle without inertia (overdamped) can be viewed as a passive inertial particle in a modified potential landscape as determined by the Hamiltonian, a feature that has also been highlighted for active particles in vortical fluid flows~\cite{2310.01829}.
Additionally, we extended the spheroidal active particle model to oblate spheroidal particles. 
We also provided the explicit form of an additional constant of motion which exists in cases where the flow $w(x,y)$ possesses rotational symmetry. 

We provide a brief comparison between the behaviour of active particles of spherical, prolate and oblate spheroidal shapes in \cref{sec:comparisons}.
Apart from some specific differences highlighted therein, one might broadly conclude that, for an infinitesimally small self-propelled particle moving with a constant speed along an axis which rotates as a result of shear in the fluid flow, the shape of the particle is not overly important.
Certainly, the same types of motion are generally observed in our calculations using an approximation of square duct flow, especially between prolate spheroidal and spherical particles.
Oblate spheroidal particles are similar to a lesser extent, a key difference being an affinity for diagonally aligned trajectories in square duct flow.

There are various directions in which this active particle model could be extended, e.g. general ellipsoidal particle shapes, time dependent flows, modelling particle collision with duct walls, particles of finite size (and interacting with the flow), and/or more complex flows such as those arising from curved duct geometry.
However, it is unclear whether such extensions may continue to possess a Hamiltonian formulation. 
It could be insightful to undertake a thorough exploration of the dynamics of the prolate and oblate spheroidal active particles within a square duct geometry, i.e. akin to \cite{ValaniHardingStokes2024}, although we generally expect the qualitative differences compared to a spherical particle to be subtle.

\appendix
\section{Verification of the Hamiltonian in the case of a prolate spheroidal particle}\label{app:verification}

We derived a Hamiltonian for the motion of an active prolate spheroidal particle primarily by considering the momenta $P_x$ and its time derivative.
Here we verify that the Hamiltonian is correct by illustrating that we can extract the correct equation governing $\ddot{y}$ via the application of Hamilton's equations to the momenta $P_y$.

The corresponding Hamilton equation is
\begin{align*}
\dot{P}_y = -\frac{\partial\mathcal{H}}{\partial y} 
&=\frac{P_x^2+P_y^2}{2F_p(w)^2}F_p^\prime(w)w_y-\frac{1}{2G}F_p^\prime(w)w_y\left[\frac{1-G}{2}+\frac{1+G}{2F_p(w)^2}\right] \\
&=\frac{1-e_z^2}{2}Ge_zF_p(w)w_y-\frac{1}{2}e_zF_p(w)w_y\left[\frac{1-G}{2}+\frac{1+G}{2F_p(w)^2}\right] \,,
\end{align*}
which will be equated with 
\begin{align*}
\dot{P}_y =\frac{d}{dt}[\dot{y}F_p(w)]
&=\ddot{y}F_p(w)+\dot{y}F_p^\prime(w)(\dot{x}w_x+\dot{y}w_y) \\
&=\ddot{y}F_p(w)+Ge_ye_zF_p(w)(e_xw_x+e_yw_y) \,.
\end{align*}
After equating the two and re-arranging with $\ddot{y}$ on one side we find
\begin{align*}
\ddot{y}
&=\frac{1-e_z^2}{2}Ge_zw_y-\frac{1}{2}e_zw_y\left[\frac{1-G}{2}+\frac{1+G}{2F_p(w)^2}\right]-Ge_ye_z(e_xw_x+e_yw_y) \\
&=-\frac{F_p(w)^2-1}{2F_p(w)^2}\frac{G}{f_p(G)^2}e_zw_y-\frac{1}{2}e_zw_y\left[\frac{1-3G}{2}+\frac{1+G}{2F_p(w)^2}\right]-Ge_ye_z(e_xw_x+e_yw_y) \\
&=-\frac{1}{2}e_zw_y\left[\frac{1-3G}{2}+\frac{1+G}{2F_p(w)^2}+\frac{G}{f_p(G)^2}-\frac{G}{f_p(G)^2F_p(w)^2}\right]-Ge_ye_z(e_xw_x+e_yw_y) \\
&=-\frac{1}{2}e_zw_y\left[\frac{1-3G}{2}+\frac{1+G}{2F_p(w)^2}+\frac{1+G}{2}-\frac{1+G}{2F_p(w)^2}\right]-Ge_ye_z(e_xw_x+e_yw_y) \\
&=-\frac{1}{2}e_zw_y(1-G)-Ge_ye_z(e_xw_x+e_yw_y) \,,
\end{align*}
as desired.

Lastly, we verify that the Hamiltonian is a constant of motion for the system.
For convenience, we first express $\mathcal{H}$ entirely in terms of $F_p(w)$, that is
\begin{align*}
\mathcal{H}
&=\frac{1-e_z^2}{2}F_p(w)+\frac{1}{2G}\left[\frac{1-G}{2}F_p(w)-\frac{1+G}{2F_p(w)}\right]+\frac{1}{2} \\
&=\frac{1}{2}F_p(w)-\frac{F_p(w)^2-1}{2f_p(G)^2F_p(w)}+\frac{1}{2G}\left[\frac{1-G}{2}F_p(w)-\frac{1+G}{2F_p(w)}\right]+\frac{1}{2} \,.
\end{align*}
Now observe that
\begin{align*}
\dot{\mathcal{H}} &= \frac{dw}{dt}F_p^\prime(w)\left(\frac{1}{2}-\frac{F_p(w)^2+1}{2f_p(G)^2F_p(w)^2}+\frac{1}{2G}\left[\frac{1-G}{2}+\frac{1+G}{2F_p(w)^2}\right]\right) \\
&=\frac{dw}{dt}F_p^\prime(w)\left(\frac{1}{2}-\frac{1}{2f_p(G)^2}-\frac{1}{2f_p(G)^2F_p(w)^2}+\frac{1-G}{4G}+\frac{1+G}{4GF_p(w)^2}\right) \\
&=\frac{dw}{dt}F_p^\prime(w)\left(\frac{1}{2}-\frac{1+G}{4G}-\frac{1+G}{4GF_p(w)^2}+\frac{1-G}{4G}+\frac{1+G}{4GF_p(w)^2}\right) \\
&=\frac{dw}{dt}F_p^\prime(w)\left(0\right)=0 \,.
\end{align*}
Thus the Hamiltonian is indeed a constant of motion. 

\section{Bounds on the potential \texorpdfstring{$V_p$}{Vp}}\label{app:potential_properties}

We claim in the main text that $V_p\in[0,1/2]$ throughout the motion of any active prolate spheroidal particle (and similarly for $V_o$ in the case of oblate spheroidal particles).
To show this, we first observe that $V_p(w)$ may be expressed as
\begin{align*}
V_p(w(x,y)) &= \frac{F_p(w)}{2G}\left[\frac{1-G}{2}-\frac{1+G}{2F_p(w)^2}\right]+\frac{1}{2} \\
&= \frac{F_p(w)}{2G}\left[\frac{1}{2}\frac{F_p(w)^2-1}{F_p(w)^2}-\frac{G}{2}\frac{F_p(w)^2+1}{F_p(w)^2}\right]+\frac{1}{2} \\
&= \frac{F_p(w)}{2G}\left[\frac{1}{2}f(G)^2e_z^2-\frac{G}{2}(2-f_p(G)^2e_z^2)\right]+\frac{1}{2} \\
&= \frac{F_p(w)}{2}\left[\frac{1+G}{2G}f_p(G)^2e_z^2-1\right]+\frac{1}{2} \\
&= \frac{F_p(w)}{2}\left[e_z^2-1\right]+\frac{1}{2} \,,
\end{align*}
noting we have utilised the fact that 
\begin{equation*}
f_p(G)^2e_z^2=\tanh\left(C_p+\frac{1+G}{2}f_p(G)w\right)^2=\frac{F_p(w)^2-1}{F_p(w)^2} \,.
\end{equation*}
Rearranging this latter identity (and noting $F_p(w)>0$) leads to
\begin{equation*}
F_p(w)=\frac{1}{\sqrt{1-f_p(G)^2e_z^2}}
\end{equation*}
which may be utilised to express $V_p$ entirely in terms of $e_z$, that is
\begin{equation*}
V_p = \frac{1}{2}\left(\frac{e_z^2-1}{\sqrt{1-f_p(G)^2e_z^2}}+1\right) \,.
\end{equation*}
Now $e_z^2\in[0,1]$, $V_p|_{e_z^2=0}=0$, $V_p|_{e_z^2=1}=\frac{1}{2}$ and $V_p$ is an increasing function of $e_z^2$; also $0<G<1$ so that $0<f_p(G)<1$ and $1-f_p(G)^2e_z^2>0$. 
We therefore conclude that indeed $V_p\in[0,1/2]$ for every orbit.

%\section*{Acknowledgments}
%We would like to acknowledge the assistance of volunteers in putting
%together this example manuscript and supplement.

\bibliographystyle{siamplain}
\bibliography{references}

\end{document}